# Mass-transport-limited reaction rates and molecular diffusion in the van der Waals gap beneath graphene

Hossein Mirdamadi[1#], Jiří David[1,2#], Rui Wang[3], Tianle Jiang[4], Yanming Wang[5], Karel Vařeka[2], Michal Dymáček[2], Petr Bábor[1,2], Tomáš Šikola[1,2], Miroslav Kolíbal[1,2*]

[1]CEITEC BUT, Brno University of Technology, Purkyňova 123, 612 00 Brno, Czech Republic

[2]Institute of Physical Engineering, Brno University of Technology, Technická 2, 616 69 Brno, Czech Republic

[3]School of Mechanical Engineering, Shanghai Jiao Tong University, 800 Dongchuan Rd, Shanghai, China, 200240

[4]University of Michigan-Shanghai Jiao Tong University Joint Institute, Shanghai Jiao Tong University, 800 Dongchuan Rd, Shanghai, China, 200240

[5]Global Institute of Future Technology, Shanghai Jiao Tong University, 800 Dongchuan Rd, Shanghai, China, 200240

[#]contributed equally to this work

[*]kolibal.m@fme.vutbr.cz

**Abstract**

The confinement of molecules within the van der Waals (vdW) gap between a two-dimensional 2D material and a catalytic substrate offers a promising route toward the development of molecule-selective catalysts with increased reaction rates. However, identifying the kinetic limitations of such confined reactions remains challenging. Here, we employ an inverted wedding-cake configuration of multilayer graphene on platinum to study the dynamics of graphene etching in the vdW gap by various molecules ($O_2$, $H_2$, and CO), using in situ scanning electron microscopy. Under the experimental conditions explored (up to $p = 1.4\times10^{-3}$ Pa and $T = 1000$ °C), the etching reaction rates are limited by mass transport within the confined space. This limitation persists even for CO, despite its anomalously enhanced transport resulting from a significant lifting of the vdW gap. Reactive molecular dynamics simulations further reveal multiple etching pathways for CO, enabled by confinement within the vdW space. Once mass-transport limitations are overcome, the vdW gap acts as an effective nanoreactor, facilitating reaction pathways that would be otherwise inaccessible on a pristine surface without spatial confinement.

**Introduction**

Intercalation is the insertion of foreign atoms, molecules, and ions within the matrix of layered materials. The most successful technology that builds on this phenomenon is lithium-ion battery, derived from the initial research of Stanley M. Whittingham, focused on the intercalation of lithium ions within layered $TiS_2$ [1]. In a follow-up studies, graphite and oxide electrodes have been introduced [2–5], facilitating the rapid developments of lithium battery technology. Recently, with renewed interest in 2D materials, the intercalation of foreign atoms in the van der Waals space between individual layers and/or the substrate has been utilized for many purposes [6]. After intercalation, the distance between the layers may increase; an intercalated crystal could be utilized for gas storage[6] and, in addition, the intercalated layers are easier to peel off due to weakened van der Waals bonding[7]. In the case of monolayer 2D materials positioned on the substrate surface, weakening of the interaction between the 2D material and the substrate due to intercalation may uncover the inherent electronic structure of the 2D material[6], [8], bringing the original promise of unrivalled electronic properties and applications in electronic devices back on stage.[9] Moreover, intercalated 2D materials have been reported to exhibit properties unseen in their free-standing counterparts.[10,11]

While many researchers focus on the properties of the 2D materials after intercalation, it is exciting to study the effect of space confinement within the van der Waals gap in between the substrate and a 2D material cover.[12,13] The entry into the gap is either from edges or, potentially, through defects within the 2D material. [14, 15,16] Such asymmetric entrance pathway facilitates, e.g., the formation of new 2D materials within the gap, as has been demonstrated for 2D GaN [17], air-sensitive monolayer indium [16,18], and others[19]. Another appealing research direction is performing chemical reactions within the van der Waals gap[20–26], in particular gas-phase heterogenous catalysis [27]. Several recent reports suggest that within the gap, the catalytic activity of the substrate that drives the reaction is increased [25], potentially due to lowered activation energies of related temperature-activated processes[28,29] via charge-transfer from 2D monolayer [30]. Theoretical modelling suggests significantly increased pressure within the gap [13,20] which may affect molecular shapes and, hence, reaction rate. Enhancement can also come from the third species present during the reaction, facilitating the entrance of the van der Waals gap for larger molecules involved in reaction of interest. [31]

On metal surfaces, in the absence of molecular species, the graphene edges are bonded to the metal atoms, effectively sealing the space between the metal and the graphene [32,33]. The intercalation is possible only after the edges become terminated by the adsorbates and "lifted" up, thus opening the van der Waals space for diffusing atoms and molecules.[33] It has been previously shown that this process is viable at very low temperatures[33–35] and, thus, even the prolonged exposure of a graphene-covered metal to the atmosphere may result in intercalation[36]. Hence, a mechanistic understanding of the initiation of the intercalation process has been established. Direct microscopic observations of diffusion are possible via transmission electron microscopy[13], yet, these experiments are limited to a specific case of atomic motion within 2D material stacks. Quantitative experimental data on atomic or molecular dynamics in the van der Waals gap between a metal and 2D material are missing. Only recently have such data been reported via indirect, yet ingenious experimental observations. [31,37,38, 39]

Going even one step further, the full understanding of the chemical reaction kinetics requires the knowledge of limiting processes, reaction rates, diffusion coefficients, etc. Here, we utilize in situ electron microscopy and perform etching of multilayer graphene in the inverse wedding cake configuration prepared on top of the platinum surface. We have chosen to investigate $O_2$, $H_2$ and CO due to their principal importance in fundamental catalytic reactions (e.g., CO oxidation over platinum). We show that for all three gas molecules of interest, the etch rate under the graphene is mass transport-limited (within the pressure range used here; up to $1.4\times10^{-3}$ Pa) and that the etching rate of the buried graphene layer is smaller compared to the etching rate of the top-layer graphene. Moreover, we have derived the activation energies for diffusion and compared them to the diffusion on the free surface. The enhanced etching reaction rates within the van der Waals gap, as reported by others for different reactions as well [40], are not observed for reaction conditions employed here. Nevertheless, molecular dynamics simulations suggest different alternative pathways for graphene etching by CO, which are limited to the van der Waals space and are not observed on the pristine platinum surface.

## Results

**Graphene intercalation**. Direct microscopic observation of the intercalation is quite challenging. Instead, one can utilize the secondary effects caused by intercalation to indirectly monitor the process. X-Ray Photoelectron Spectroscopy (XPS) is commonly used to monitor intercalation of 2D materials, because the change in interlayer coupling after intercalation results in different binding energies[33,41]. However, because of weak coupling of graphene to Pt, the shifts of the relevant binding energies are too small to provide a conclusive proof of intercalation in our experiments (see Fig. S1, Supporting Information). As an alternative to XPS, observation of moiré pattern fading has been employed before[42,43]. Moiré effects appear in both scanning tunneling microscopy[24] as well as in electron diffraction[42,43]. The disappearance of the moiré pattern upon intercalation has previously been related to the weakened coupling between the layers. However, here we show that specifically in the case of electron diffraction, the moiré pattern intensity in the diffraction image is rather a measure of the vdW gap height and cannot be used solely as an indicator of intercalation. Hence, in order to get detailed insight into the intercalation process, we have employed Low Energy Electron

Diffraction (Fig. 1a,c) together with measurement of electron reflectivity (Fig. 1d) in Low Energy Electron Microscope (LEEM) on a single crystal (111) platinum partially covered with monolayer graphene (see Scanning Electron Microscope (SEM) image in Fig. 1c). Similar to other reports,[33] we were not able to intercalate graphene if a full monolayer was prepared on the metal surface.

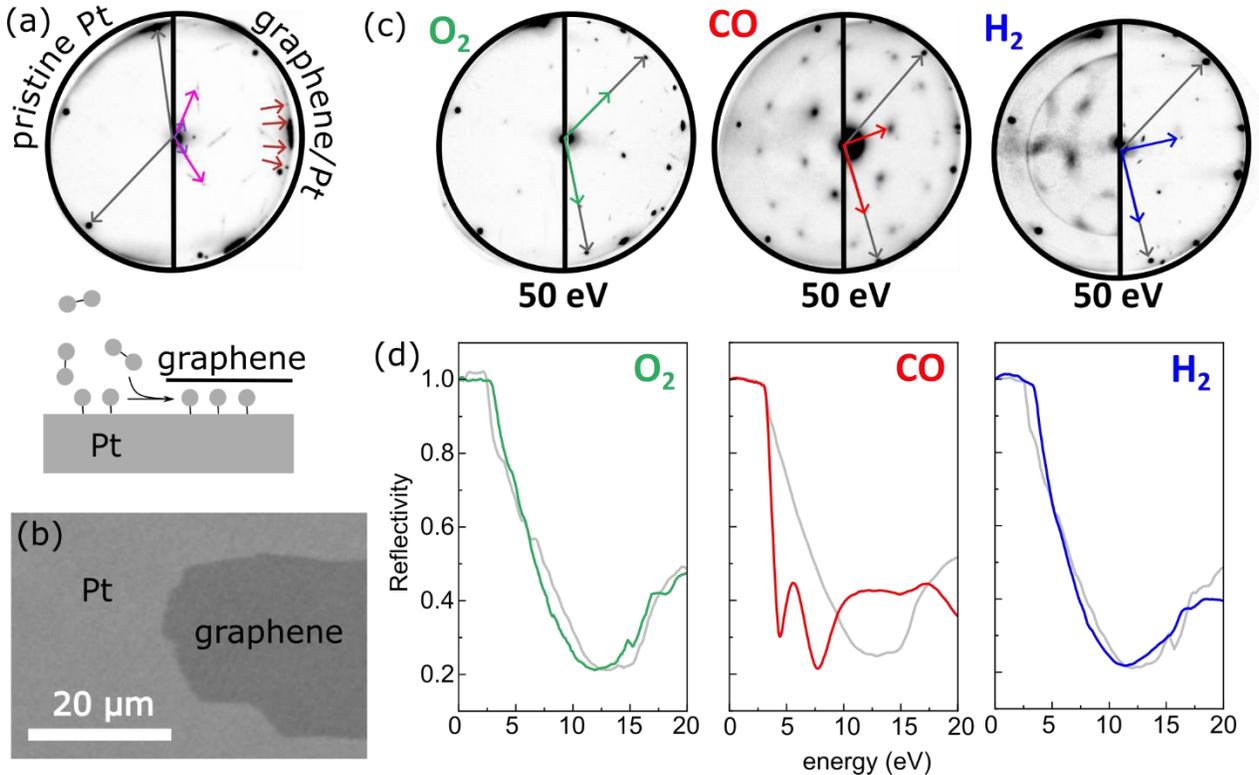

Fig.1: Intercalation of different molecules under the graphene, documented by low-energy electron diffraction (LEED, panel c) and reflectivity of electrons (d). (a) Low-energy electron diffraction patterns (50 eV), showing pristine Pt(111) (left) and graphene-covered Pt(111) (right) – see schematic at the bottom. Several of the many graphene-related spots at the circumference of the pattern are marked by maroon arrows. The other arrows depict the unit vectors of the respective unit cells: Pt(111) (grey), graphene moiré demonstrated by the spots around the central (0,0) spot (violet), $(\sqrt{7} \times \sqrt{7})R19.1°$ graphene superstructure (pink). (b) SEM image of the monolayer graphene flake. (c) LEED patterns taken at pristine Pt(111) (left) and graphene-covered platinum (right) after exposure to $O_2$, CO and $H_2$ at room temperature. CO diffraction patterns were obtained as a composition of diffraction patterns taken at energies of 10-50 eV. The arrows represent the unit vectors of the respective unit cells. Patterns on graphene-covered platinum (right) appeared after a prolonged period compared to that of pristine platinum. These times differ for each molecule (see the text for details). (d) Reflectivity-energy curves acquired by sweeping the incident energy of electrons in LEEM. Gray curves: platinum covered with graphene under vacuum conditions; colored curves: platinum covered with graphene after exposure to the respective gas ($1–2×10^{-6}$ Pa) at different times (up to 90 min for $O_2$, 10-20 min for CO, 20-30 min for $H_2$) at room temperature.

First, exposure of the sample to different molecular gases was performed in LEEM (see Methods) at room temperature. The diffractograms acquired using the microdiffraction aperture on pristine Pt changed quickly in response to gas exposure (within 2-3 minutes), reflecting the formation of adsorbate superstructures at the surface (Fig. 1c, left half circle). Exposure to molecular oxygen results into (2×2)-O superstructure formation, while both CO and $H_2$ induce c(4×2) superstructure, in agreement with previous reports ([$O_2$[44], CO [45,46]). At areas covered with graphene, the appearance of new spots was significantly delayed (up to 90 min for $O_2$, 10-20 min for CO, 20-30 min for $H_2$). In all cases, the gas-induced superstructures were the same at pristine and graphene-covered platinum, suggesting that confinement of the adsorbate molecules in the van der Waals space does not affect preferred adsorption sites irrespective of the graphene cover [33,47]. An important difference is that the moiré-related spots, which are visible on graphene-covered areas, disappear only after intercalation by CO. O- and H-intercalated graphene areas still exhibit moiré-related spots in the diffractograms, although with lower intensity.

Reflection of low-energy electrons from the surface is a very sensitive measure of any surface modification[48]. The energy dependence of the reflectivity measured in graphene-covered areas shows a characteristic decrease at around 3-5 eV (the work function of pristine Pt is ~ 5.8 eV[49]), profound minimum between 12-15 eV, followed by a local maximum peaking at 20 eV (Fig. 1d). Intercalation of $H_2$ and $O_2$ results in a very small shift of the reflectivity minimum towards lower energies. In contrast, CO intercalation induces dramatic changes in the reflectivity: The shift to the lower energies is relatively large and an additional minimum shows up. It has been shown previously that the minimum in the reflectivity shifts toward lower energies because of the increase of the van der Waals gap height. In case of a high vdW height, additional minima may appear [50]. Compared to $O_2$ and $H_2$ intercalation, our reflectivity data thus suggest that the CO intercalation significantly increases the vdW gap height. Importantly, the gap remains lifted even at elevated temperatures (see SI, Fig. S2) while, at the same time, the adsorbate-induced diffraction patterns from below graphene disappear. This observation is important for deducing the mass transport mechanism of intercalated molecules, and this will be discussed later.

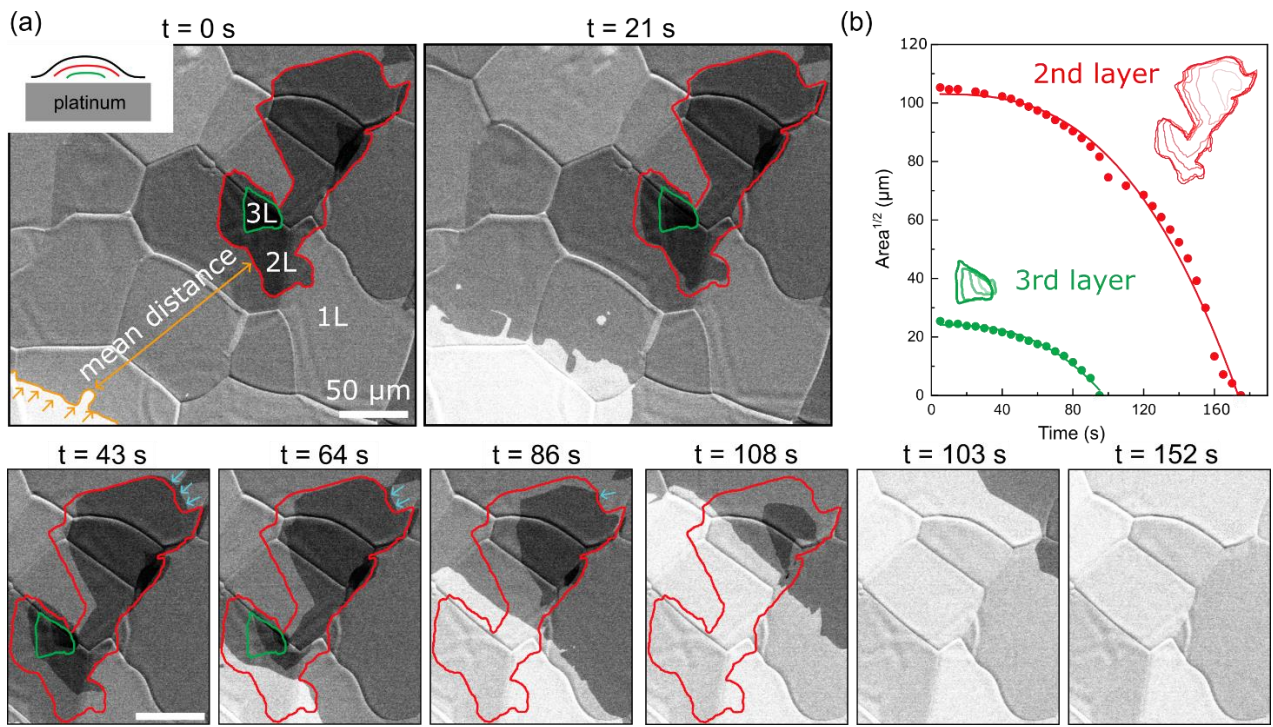

Fig. 2: Image sequence of multilayer graphene etching by oxygen observed in situ in SEM (5 keV electron beam, beam current 3 nA). (a) The etching front of the first layer (the direction of the etching front movement is marked with orange arrows) moves toward the multilayer region. Both the second (red) and third (green) buried graphene layers start to etch even before the overlayer etch front reaches their edges. Additionally, graphene growth was observed on the opposite side of the flakes for a short period (blue arrows). Oxygen pressure was $6.6 \times 10^{-4}$ Pa, temperature 1000 °C. The inset shows a multilayer graphene stack (black, overlayer; green, red, buried layers). (b) Evolution of the square root of the area of the second and third layers over time during etching for the flakes shown in (a).

**Kinetics of buried graphene etching in the vdW space**. We have followed the kinetics of intercalated molecules in the van der Waals gap indirectly by monitoring the etching of the buried graphene layers of a multilayer graphene. First, we have prepared multilayer graphene in inverted weeding cake configuration (schematically shown as inset in Fig. 2, see Methods for description). Still at elevated temperature, we introduced a precursor for etching ($O_2$, $H_2$ or $CO$) while monitoring the multilayer flakes in real time with an electron beam. The experimental conditions (up to p = $1.4 \times 10^{-3}$ Pa and T = 1000 °C) were chosen to avoid etching outbursts within the basal plane of the graphene overlayer; instead, we managed to form an etch front that moves across a sample in one direction (see Fig. 2a, orange arrows). As the etch front advances, the buried graphene layers start to etch as well. Note that the etching of the buried layers starts only after the etch front approaches them at a certain distance. Far away from the overlayer etch front, no etching of buried graphene layers is observed. We will call the distance between the overlayer etch front and the buried graphene edge 'mean distance' in the following text, since it is apparent that the etching is initiated by the precursor molecules that diffused towards the buried graphene layers within the van der Waals gap after intercalation under the overlayer graphene.

The electron beam permits visualization of the reaction and monitoring its progress in real time and space, with negligible effect on the reaction rate (see Fig. S3, Supporting Information). Therefore, we can quantitatively evaluate the scaling of the etching reaction kinetics, which provides identification of the limiting mechanism in play. If the reaction is limited by detachment of the reaction products (the so-called detachment-limited regime), the area etch rate $\frac{dA}{dt}$ scales with perimeter $r$. Therefore, the radius of the flake decreases linearly with time ($r = r_0 - \varepsilon . t$, where $\varepsilon$ is the (constant) etch rate). In case the reaction is limited by insufficient feeding with etchant molecules (mass-transport limited regime), the reaction kinetics is non-linear, $r = r_0 - \varepsilon . t^x$, where $x > 1$. In Fig. 2b, we have plotted the evolution of square root of area of the flake in time for the second and third graphene layers in the multilayer stack (Fig. 2a). The non-linearity of both curves ($x \sim 2.8$) thus suggests that graphene etching proceeds in the mass-transport-limited regime. To ensure that the non-linearity is not caused by the precursor flow anisotropy perpendicular to the etching front, we have inspected the etching of much smaller flakes than those shown in Fig. 2a (with $r \ll$ mean distance; the smallest flakes were 6×6 μm$^2$). We have arrived at non-linearity similar to the larger flakes. The anisotropy of the precursor flux thus only increases the non-linearity of the $r$-dependence, but is not solely causing it. However, the flux anisotropy is present, which is manifested by the faster recession close to the etch front (Fig. 2b), as well as by the growth of graphene at the other side of the flake, where the concentration of the etchant molecules is lower (see blue arrows in Fig. 2a, $t$ = 43-86 s). The latter phenomenon is rather surprising but can be easily explained by the presence of etch reaction products: carbon atoms. Unlike in on-surface reactions at the gas-solid interface, the reaction products do not easily escape from the reaction site in the vdW space, despite the fact that they could be volatile. Thus, in the absence of escape paths due to space confinement, they remain present in the van der Waals gap, possibly initiating new growth at the graphene edges, where locally high supersaturation with carbon is reached. The asymmetry of etching precursor flow in our experiment (from the etch front toward buried graphene) dictates that these would be the edges far apart from the etch front. In the case of our experiment (and also in work by Wang et al. [40]), the growth at far side of the flake is quickly reversed to etching once the local concentration of etch precursor molecules increases as the etch front progresses in time.

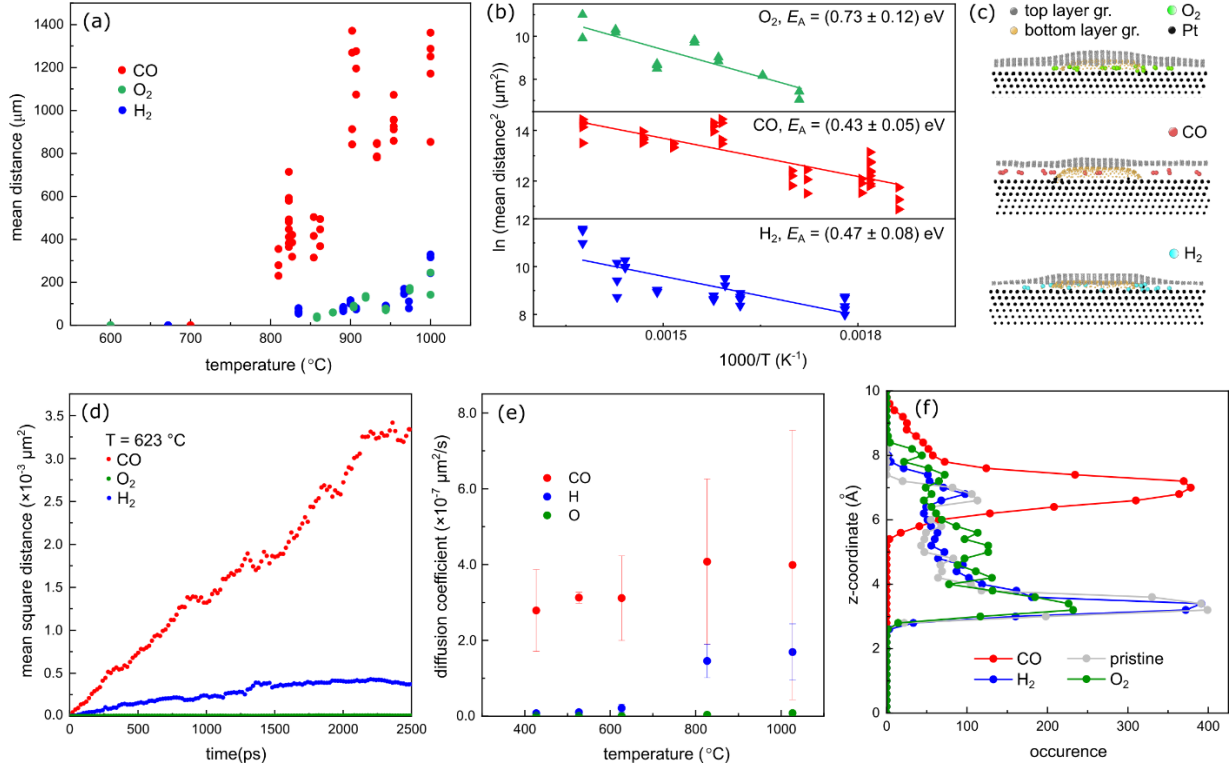

Fig. 3: (a) Dependencies of the apparent mean distance of $O_2$, $H_2$ and CO under the graphene cover on polycrystalline Pt on temperature. Data were collected for pressure range ⟨7.4×10$^{-4}$ -- 1.4×10$^{-3}$ Pa⟩. (b) Arrhenius plots (logarithm of squared mean distance dependence on inverse temperature) derived from the data in (a). (c) Molecular dynamics simulation of intercalated bilayer graphene (T = 423 °C). (d) Time-dependent mean square distance (MSD) of CO, $H_2$ and $O_2$ obtained from molecular dynamics simulation of diffusion with the vdW gap at T = 423 °C. The dependence of the diffusion coefficient on the temperature derived from the simulation is shown in (e). Panel (f) Z coordinate (height distribution) of atoms in the top graphene layer for relaxed graphene/Pt slabs shown in (c). The zero value is set to the first-layer Pt plane.

In Fig. 3a, we have plotted the mean distance derived from in situ observations of the etching (e.g., Fig. 2a) as a function of temperature. The experiments were carried out at different pressures, but the data show no correlation between the pressure and the mean distance (see Fig. S4 for details). This finding further confirms that the etching reaction in the vdW gap is mass-transport-limited.

We first assume surface diffusion to be the major mechanism of mass transport in the vdW gap. The diffusion of adatoms on a surface is commonly described as mean random walk in terms of the mean squared displacement $\langle x^2 \rangle$, in 2D case defined as $\langle x^2 \rangle = 4D\tau$, where $D$ is the diffusion coefficient and $\tau$ is the adatom residence time on the surface. Diffusion is a thermally activated process; hence, the diffusion coefficient follows an Arrhenius-type dependence on temperature, $D = D_0 \exp\left(-\frac{E_A}{k_B T}\right)$. Here, $D_0$ is the diffusion prefactor, $E_A$ the activation energy, $k_B$ is the Bolzmann constant, and $T$ is the temperature. Our measurement method does not allow us to determine the diffusion coefficient $D$ or the prefactor $D_0$ directly, because the continuously moving etching front does not permit us to measure the timing of random walks, indirectly observed in SEM. Nevertheless, the Arrhenius analysis of the mean distances in Fig. 3a provides a comparison of the activation energies for diffusion in the van der Waals space for the three molecules under study (Fig. 3b) in the mass transport-limited regime. The large scatter in the data is caused by the polycrystalline nature of the sample (the activation energies on different facets with distinct roughness can vary up to a factor 3 [51,52]). The activation energies of diffusion within the van der Waals space (0.47 eV for hydrogen, 0.43 eV for CO and 0.70 eV for oxygen) fall within the ranges reported on free platinum surfaces (0.30--0.52 eV for hydrogen [53], 0.39--0.56 eV for CO[52]), with the exception of oxygen, where the literature data oscillate between 0.43--0.58 eV[51] and 1.30--1.70 eV[52]). Hence, our data do not suggest a significant lowering of the energy barrier for diffusion in the van der Waals space. On the other hand, the CO molecule can apparently travel much larger distances (Fig. 3a) compared to H despite a similar activation energy.

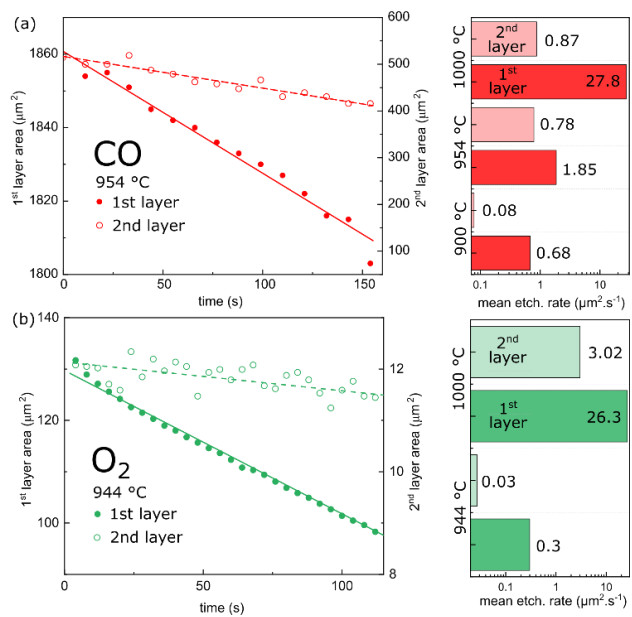

Fig. 4: Etching rates of the first (overlayer) vs. second (buried) graphene layers. Area reduction of the first and second graphene layers versus time during exposure to (a) carbon monoxide at 900 °C ($1.7\times10^{-3}$ Pa), 954 °C ($1.5\times10^{-3}$ Pa) and 1000 °C ($1.4\times10^{-3}$ Pa) and (b) oxygen at 944 °C ($5.8\times10^{-4}$ Pa) and 1000 °C ($6.6\times10^{-4}$ Pa).

In order to explain the anomalous diffusivity of CO, we have performed molecular dynamics simulation of the intercalated graphene/Pt system (see Fig. 3c for the simulation cell). By tracking individual molecules, we have been able to plot the mean square distance (plotted in Fig. 3d) and diffusion coefficient (Fig. 3e), which qualitatively resemble the experimental results: the mass transport characteristics of CO significantly exceed those of H or O. The equilibrated graphene/Pt slab after intercalation provides a clue: intercalated CO is very effective in decoupling the graphene from the Pt substrate, which is manifested by an increase in the vdW gap height (Fig. 3f), as observed experimentally in the electron reflectivity data (Fig. 1d).

Finally, we compared the etching rates of the graphene overlayer and buried graphene layer(s). In Fig. 4, we present the data for $O_2$ and CO, the two molecules relevant for the CO oxidation reaction over Pt (for $H_2$ see Supporting Information, Fig. S5). To allow for direct comparison and avoid geometrical effects (see Fig. 2b), we have extracted the etch rates from smaller areas close to the etch front (see supporting information, Fig. S6). For all the etching precursor molecules inspected, we have not observed enhancement of the etching rate of the buried layers. Instead, under the experimental conditions studied here (maximum pressure $5\times10^{-3}$ Pa, temperature up to 1000 °C), the etch rate of buried graphene is most often an order of magnitude lower than the graphene overlayer etch rate. Interestingly, it should be noted that graphene etching by CO is observed for partial CO pressures in $10^{-3}$ Pa range; At lower pressures, graphene is not etched in the CO environment (Fig. S7).

**Discussion**

The LEED and LEEM data shown in Fig.1 agree with previous reports in several aspects. Intercalation is possible even at room temperature through the edges of the graphene. The adsorption sites at both pristine Pt and under graphene cover remain the same (Fig. 1)[47]. Our data also show that the disappearance of diffraction moiré pattern is an inconclusive indicator of intercalation, as it rather reflects increase of the vdW gap height. In our work, the gap enlargement is demonstrated in case of CO molecule intercalation in electron reflectivity measurements (Fig. 1d) and further corroborated by molecular dynamics modelling (Fig. 3d). The fast CO transport observed in our experiments (Fig. 3a) and the modelling (Fig. 3d,e) is correlated with the increased vdW gap height. The weaker the confinement and the larger the vdW gap, the smaller the pressure exerted on the adsorbate molecule[13] and the less strain is imparted on the graphene [38]. This reasoning is in line with our experiments and modelling and supports the hypothesis that the faster CO transport is facilitated by another cause than the decrease in the activation energy for diffusion. An increase in diffusion prefactor (hopping frequency) could explain the anomalous data. Alternatively, it has previously been previously proposed Sutter et al.[24] and Li et al.[54] that under certain experimental conditions the increased vdW gap height may allow for fast gas-phase transport of the molecules to the reaction front, avoiding slower surface diffusion of the adatoms. Our CO-related experimental data support this scenario: at elevated temperatures, we have observed disappearance of the adsorbate-induced diffraction pattern and, at the same time, lifted van der Waals space. This observation suggests that the CO molecules are still present within the gap but in the gas phase rather than adsorbed on preferential sites at the platinum surface.

CO has recently been reported as a promising precursor for graphene growth on copper[55]. It is therefore surprising to observe graphene etching on platinum, as we experience in our experiments (see, e.g., Figs. 4 and S7). We performed secondary ion mass spectroscopy (SIMS) depth profiling of the Pt sample after exposure to CO at the etching temperature to discover a significant amount of carbon dissolved inside platinum, as compared to pristine platinum sample that was cleaned in situ and not exposed to any C-containing gas (Fig. S8). Therefore, it is plausible to assume that the CO molecule may dissociate on Pt surface (as proposed in[56]). The dissociation is followed by dissolution of atomic carbon in Pt, leaving atomic oxygen available to etch graphene and, thus, explaining the observed graphene etching. However, why graphene etching by CO occurs only at higher pressures remains to be answered.

The necessary prerequisite for molecular intercalation is termination of the graphene edges by adsorbed atoms or molecules[33]. Nonterminated graphene edges are commonly attached to the metal surface, making the vdW space effectively closed, thus preventing the intercalation. Typically, in the case of $O_2$ and $H_2$, the molecules dissociate upon adsorption at the metal surface (in the case of Pt, even at room temperature). A certain surface coverage by the atomic species is needed before the graphene edges are terminated by the H or O atoms [33]. Subsequently, the graphene edges are lifted, opening the vdW gap for the diffusing molecules and adatoms [34]. Previously, decoupling of the graphene edges from the metal has been shown to be the rate-limiting step for intercalation at temperatures below 600 °C [24]. Our data suggest that at higher temperatures, the rate-limiting step becomes the mass transport below the overlayer graphene towards the reaction front. The slower etching rates of the buried graphene layers (Fig. 4) contradict a similar recent experiment [40], as well as the encouraging reports on the acceleration of reaction rates in the van der Waals gap. However, our conclusion agrees with other reports where the enhanced etch rate of the buried layers was not observed [32,57]. Limited mass-transport thus potentially outweighs lowered activation energies for certain other reactions in the vdW gap reported in the literature[30,58,59]. It should be noted, however, that a more in depth evaluation of reaction rates under different experimental conditions is required to fully understand the confinement effect on chemical reactions in the van der Waals space, even for the reactions beyond graphene etching (e.g. CO oxidation reaction rates).

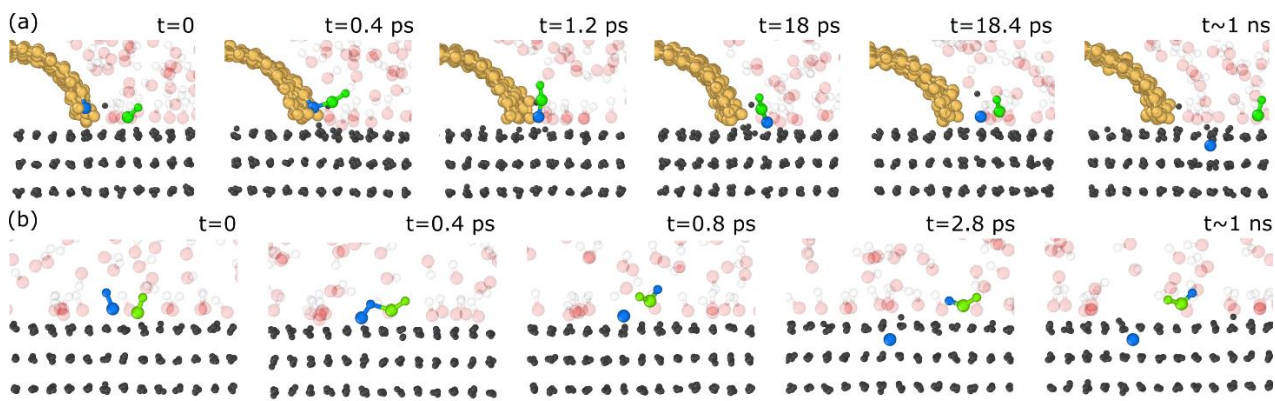

Fig. 5: Molecular dynamics simulations of CO molecules confined within two Pt surfaces with a graphene layer atop one of the Pt surfaces (the other Pt surface is not visible in the images). (a) Generation of $C_2O$ (CO+C(graphene)↔$C_2O$↔CO+C(dissolved)). (b) $CO_2$ generation (2CO→$CO_2$ + C(dissolved)). Gray: Pt atoms. Orange: graphene. Transparent red: C atom in CO molecules. Transparent white: O atom in CO molecules. Reactive atoms are colored blue and green (where (a) green is a CO molecule, (b) blue is a C atom in graphene, and (c) blue and green are CO molecules); large spheres are C atoms, small ones are O atoms. The temperature is set at 1700 K.

In case of CO-induced etching, our simulations show that within the van der Waals gap, different graphene etching reaction paths are also plausible. We have performed molecular dynamics simulations of dense CO gas confined within two platinum surfaces, including a graphene flake localized at one of these surfaces, at 1700-1750 K. Importantly, we have observed additional reaction pathways under confinement that were not present on the pristine Pt surface. For example, carbon uptake by CO via an intermediate $C_2O$ formation (Fig. 4c)[60], potentially followed by graphene etching and C dissolution ($C_2O \rightarrow CO+C(dissolved)$). Another kinetically feasible pathway is the formation of $CO_2$ (Fig. 4d), again potentially followed by graphene etching ($C(graphene)+CO_2 \rightarrow 2CO$).[61,62] The enhanced CO pressure under confinement in these MD simulations is considered realistic, given that similar high-pressure conditions have been reported in the van der Waals gap[13]. Direct experimental confirmation is, however, rather challenging given instrumental limitations of in situ techniques, despite recent progress in such observations[59].

**Conclusions**

In summary, we have utilized in situ electron microscopy for monitoring buried graphene etching within the van der Waals space between the platinum substrate and CVD-grown graphene. Our data demonstrate that the reaction is mass-transport limited for the three gases in the study ($O_2$, $H_2$, CO) within the accessible experimental range (up to $1.35 \times 10^{-3}$ Pa). We have always observed that the etching reaction is slower within the van der Waals gap. In order to utilize the van der Waals gap for more efficient catalytic reactions, mass transport within the gap has to be accelerated, e.g. by increasing the gap height or choosing another 2D material as the cover.

**Materials and Methods**

The initial experiments shown in Fig. 1 were carried out in an ultra-high vacuum (UHV) system (base pressure ~$10^{-10}$ mbar), which houses a SPECS FE-LEEM P90 low-energy electron microscope and SPECS X-ray photoelectron spectroscope (XPS) with the Mg Kα X-Ray source and Phoibos 150 spectrometer, among other tools, and allows transfer of samples *in vacuo* under UHV conditions. A single crystal sample of Pt(111) (SPL, The Netherlands) was annealed at 905 °C in oxygen atmosphere ($1.3\times10^{-4}$ Pa) to remove all residual contamination. The sample cleanliness was checked by LEED, XPS, and bright-field LEEM imaging. Graphene was grown in the preparation chamber by exposing the clean Pt sample to ethylene gas ($5-7\times10^{-6}$ Pa) at 950 °C. Molecular intercalation was performed directly in the LEEM chamber by introducing the relevant gas (Oxygen 5.0 (Messer), Hydrogen 5.0 (Messer) and CO 3.7 (Linde)) with a leak valve. The intercalation pressures were $1.6\times10^{-4}$ Pa.

The sample utilized for all etching experiments was Pt wire (0.2 mm or Goodfellow, 0.25 mm diameter, 99.99% purity), which was hammered to expand the planar surface for graphene growth. Subsequently, the sample was annealed at an elevated temperature (> 1000 °C) in the presence of oxygen inside an ultra-high vacuum chamber equipped with scanning electron microscope. The base pressure of the main chamber reaches $4.8\times10^{-7}$ Pa. The sample annealing was done by direct current resistive heating and was performed until all detectable carbon residues were removed and a clean surface with well-defined grains was achieved. Temperature was measured with an infrared pyrometer Micro-Epsilon CT-M3 H1-SF at a wavelength of 2.3 μm. The emissivity was set to 0.1. To grow graphene, ethylene gas (Messer, Ethylene 3.5) was supplied to the microscope chamber by a leak valve. First, we grew a monolayer of graphene through ethylene decomposition at $1-2\times10^{-4}$ Pa at elevated temperatures (in the range of 900 °C to 1000 °C). Then, to form an inverted wedding cake configuration, we dissolved the graphene in the platinum substrate by increasing the substrate temperature by 100 °C with respect to the growth temperature of the first graphene layer. Subsequently, we slowly decreased the temperature, which resulted in carbon segregation and graphene flake(s) formation on the Pt surface. In addition to single-layer graphene flakes, carbon segregation leads to formation of additional graphene layers below the already segregated graphene flakes (see Fig. S9 for SEM images taken during the growth process). Such multilayer configuration is called an inverted wedding cake. To perform etching, the etching agents were introduced into the chamber by additional leak valves.

Scanning electron microscopy observation during the etching reaction was performed using a 5 keV electron beam with a current of 3 nA. Secondary electron images were formed by the Everhart-Thornley detector inside the main chamber.

In situ XPS analysis was performed without charge neutralization. Survey spectra were collected in a high magnification mode using a pass energy of 100 eV with a dwell time of 100 ms and a 1 eV energy step. Detailed spectra were acquired in a high-magnification mode using a pass energy of 25 eV with the 500 ms dwell time and 0.1 eV energy step, integrating 20-30 sweeps. All the spectra were collected in a normal emission geometry (emission angle parallel to the surface normal). No spectral shifts were performed. The spectra were further processed using Casa XPS software. The C 1s, O 1s and Pt $4p_{3/2}$ peaks (Fig. S1) were fitted using the Shirley background and several components described in the main text and in the table below: BE, binding energy; shift, with respect to the first component; FWHM, full width at half-maximum.

| C 1s | shape | BE (eV) | shift (eV) | FWHM (eV) |
|---|---|---|---|---|
| Graphene | assymetric Voigt: LA(1.53,243) | 283.9 | - | 1.06 |
| Graphene intercalated | assymetric Voigt: LA(1.53,243) | 283.8 | 0.1 | 1.10 |
| CO on Pt (111) | assymetric Voigt: LA(1.53,243) | 286.4 | - | 1.87 |
| CO beneath graphene | assymetric Voigt: LA(1.53,243) | 286.2 | 0.2 | 2.05 |

| O 1s | shape | BE (eV) | shift (eV) | FWHM (eV) |
|---|---|---|---|---|
| CO on Pt (111) | assymetric Voigt: LA(1.45,150) | 532.08 | - | 3.47 |
| CO beneath graphene | assymetric Voigt: LA(1.45,150) | 531.71 | 0.37 | 2.60 |

| Pt $4p_{3/2}$ | shape | BE (eV) | shift (eV) | FWHM (eV) |
|---|---|---|---|---|
| Pt + ads. CO | assymetric Voigt: LA(2.03,4.7,0) | 519.23 | - | 5.01 |
| Pt + graphene | assymetric Voigt: LA(2.03,4.7,0) | 518.95 | -0.28 | 4.53 |
| Pt + CO intercalated graphene | assymetric Voigt: LA(2.03,4.7,0) | 519.15 | -0.08 | 4.78 |

Depth profiling of platinum samples was performed using Secondary Ion Mass Spectrometry (SIMS) to elucidate sub-surface compositional variations of clean and gas- and temperature-treated Pt samples. All analyses were conducted using a TOF-SIMS5 instrument (IONTOF GmbH, Germany), allowing for high-resolution depth profiling at the nanometer scale. $Bi^+$ primary ions were employed under the following conditions: impact energy of 30 keV, incidence angle of 45°, pulsed primary current of approximately 3 pA, and raster dimensions of 100 μm × 100 μm². To enhance depth resolution and compensate for matrix effects, $Cs^+$ (2 keV, 130 nA) co-sputtering was used in non-interlaced sputtering mode. The two beams (sputtering and analysis) were alternating as follows: two $Cs^+$ sputtering frames were followed by a 0.1s pause and by four $Bi^+$ analysis frames. Measurements were carried out in negative polarity mode with an electron flood gun to compensate for surface charging during sputtering. The resulting depth profiles were plotted as signal intensity versus sputter time. These profiles facilitated a qualitative, relative comparison of carbon distribution across differently treated platinum samples.

Classical molecular dynamic simulations were performed using the LAMMPS simulation package.[63] The reaxFF$_{C/H/O/Pt}$ interatomic potential[64] was adopted, which has been proven capable of describing complex chemical processes and reasonably accurate in modeling the interactions between carbon (C), hydrogen (H), oxygen (O), and platinum (Pt) atoms. Each MD configuration of the intercalated graphene/Pt system consisted of a single-crystalline Pt (111) substrate with a thickness of 1.2 nm and two layers of graphene (overlayer and buried layer) above in an inverse wedding cake configuration. In each system, 10 CO, $H_2$ or $O_2$ gas molecules were randomly added in the region between the graphene and the substrate, and above the graphene layers, a vacuum region was created with a thickness of around 5 nm. The *x* and *y* dimensions of the simulation boxes were maintained at 9.00 nm and 7.80 nm respectively under periodic boundary conditions, as shown in Fig. S10. A Nose−Hoover thermostat[65] was employed to control the temperature at 300, 700, 800, 900, 1100, 1300,1500 K, respectively. The velocity Verlet algorithm [66] with a time step $\Delta t = 0.2$ fs was applied and the total length of each simulation was greater than 3 ns. For each condition investigated in this study, MD runs were repeated five times independently, and the coordinates of atoms were recorded every 1 ps for postprocessing analyses.

**References**

(1) Whittingham, M. S.; Gamble, F. R. The Lithium Intercalates of The Transition Metal Dichalcogenides. *Mat. Res. Bull* **1975**, *10* (5), 363–371. https://doi.org/10.1016/0025-5408(75)90006-9.

(2) Godshall, N. A.; Raistrick, I. D.; Huggins, R. A. Thermodynamic Investigations of Ternary Lithium-Transition Metal-Oxygen Cathode Materials. *Mat. Res. Bull* **1980**, *15* (5), 561–570. https://doi.org/10.1016/0025-5408(80)90135-X.

(3) Mizushima, K.; Jones, P. C.; Wiseman, P. J.; Goodenough, J. B. Li$_x$CoO$_2$ (0<x≤l): A New Cathode Material for Batteries of High Energy Density. *Mat. Res. Bull* **1980**, *15* (6), 783–789. https://doi.org/10.1016/0025-5408(80)90012-4.

(4) Yazami, R.; Touzain, P. H. A Reversible Graphite-Lithium Negative Electrode for Electrochemical Generators. *J Power Sources* **1983**, *9* (3), 365–371. https://doi.org/10.1016/0378-7753(83)87040-2.

(5) Besenhard, J. O. The Electrochemical Preparation and Properties of Ionic Alkali Metal-and NR4-Graphite Intercalation Compounds in Organic Electrolytes. *Carbon N Y* **1976**, *14* (2), 111–115. https://doi.org/10.1016/0008-6223(76)90119-6.

(6) Daukiya, L.; Nair, M. N.; Cranney, M.; Vonau, F.; Hajjar-Garreau, S.; Aubel, D.; Simon, L. Functionalization of 2D Materials by Intercalation. *Prog Surf Sci* **2019**, *94* (1), 1–20. https://doi.org/10.1016/j.progsurf.2018.07.001.

(7) Yang, R.; Fan, Y.; Mei, L.; Shin, H. S.; Voiry, D.; Lu, Q.; Li, J.; Zeng, Z. Synthesis of Atomically Thin Sheets by the Intercalation-Based Exfoliation of Layered Materials. *Nature Synthesis* **2023**, *2* (2), 101–118. https://doi.org/10.1038/s44160-022-00232-z.

(8) Rosenzweig, P.; Karakachian, H.; Marchenko, D.; Starke, U. Surface Charge-Transfer Doping a Quantum-Confined Silver Monolayer beneath Epitaxial Graphene. *Phys Rev B* **2022**, *105* (23), 235428. https://doi.org/10.1103/PhysRevB.105.235428.

(9) Robinson, J. A.; Hollander, M.; Labella, M.; Trumbull, K. A.; Cavalero, R.; Snyder, D. W. Epitaxial Graphene Transistors: Enhancing Performance via Hydrogen Intercalation. *Nano Lett* **2011**, *11* (9), 3875–3880. https://doi.org/10.1021/nl2019855.

(10) Ichinokura, S.; Tokuda, K.; Toyoda, M.; Tanaka, K.; Saito, S.; Hirahara, T. Van Hove Singularity and Enhanced Superconductivity in Ca-Intercalated Bilayer Graphene Induced by Confinement Epitaxy. *ACS Nano* **2024**, *18* (21), 13738–13744. https://doi.org/10.1021/acsnano.4c01757.

(11) Nong, H.; Tan, J.; Sun, Y.; Zhang, R.; Gu, Y.; Wei, Q.; Wang, J.; Zhang, Y.; Wu, Q.; Zou, X.; Liu, B. Cu Intercalation-Stabilized 1T′ MoS$_2$ with Electrical Insulating Behavior. *J Am Chem Soc* **2025**, *147* (11), 9242–9249. https://doi.org/10.1021/jacs.4c14945.

(12) Li, H.; Xiao, J.; Fu, Q.; Bao, X. Confined Catalysis under Two-Dimensional Materials. *Proc Natl Acad Sci U S A* **2017**, *114* (23), 5930–5934. https://doi.org/10.1073/pnas.1701280114.

(13) Längle, M.; Mizohata, K.; Mangler, C.; Trentino, A.; Mustonen, K.; Åhlgren, E. H.; Kotakoski, J. Two-Dimensional Few-Atom Noble Gas Clusters in a Graphene Sandwich. *Nat Mater* **2024**, *23* (6), 762–767. https://doi.org/10.1038/s41563-023-01780-1.

(14) Bunch, J. S.; Verbridge, S. S.; Alden, J. S.; Van Der Zande, A. M.; Parpia, J. M.; Craighead, H. G.; McEuen, P. L. Impermeable Atomic Membranes from Graphene Sheets. *Nano Lett* **2008**, *8* (8), 2458–2462. https://doi.org/10.1021/nl801457b.

(15) Sun, P. Z.; Yang, Q.; Kuang, W. J.; Stebunov, Y. V.; Xiong, W. Q.; Yu, J.; Nair, R. R.; Katsnelson, M. I.; Yuan, S. J.; Grigorieva, I. V.; Lozada-Hidalgo, M.; Wang, F. C.; Geim, A. K. Limits on Gas Impermeability of Graphene. *Nature* **2020**, *579* (7798), 229–232. https://doi.org/10.1038/s41586-020-2070-x.

(16) Pham, V. D.; González, C.; Dappe, Y. J.; Dong, C.; Robinson, J. A.; Trampert, A.; Engel-Herbert, R. Scanning Tunneling Microscopy of Ultrathin Indium Intercalated between Graphene and SiC Using Confinement Heteroepitaxy. *Appl Phys Lett* **2024**, *125* (18), 181602. https://doi.org/10.1063/5.0223972.

(17) Al Balushi, Z. Y.; Wang, K.; Ghosh, R. K.; Vilá, R. A.; Eichfeld, S. M.; Caldwell, J. D.; Qin, X.; Lin, Y. C.; Desario, P. A.; Stone, G.; Subramanian, S.; Paul, D. F.; Wallace, R. M.; Datta, S.; Redwing, J. M.; Robinson, J. A. Two-Dimensional Gallium Nitride Realized via Graphene Encapsulation. *Nat Mater* **2016**, *15* (11), 1166–1171. https://doi.org/10.1038/nmat4742.

(18) Schmitt, C.; Erhardt, J.; Eck, P.; Schmitt, M.; Lee, K.; Keßler, P.; Wagner, T.; Spring, M.; Liu, B.; Enzner, S.; Kamp, M.; Jovic, V.; Jozwiak, C.; Bostwick, A.; Rotenberg, E.; Kim, T.; Cacho, C.; Lee, T. L.; Sangiovanni, G.; Moser, S.; Claessen, R. Achieving Environmental Stability in an Atomically Thin Quantum Spin Hall Insulator via Graphene Intercalation. *Nat Commun* **2024**, *15* (1), 1486. https://doi.org/10.1038/s41467-024-45816-9.

(19) Ryu, H.; Park, H.; Kim, J. H.; Ren, F.; Kim, J.; Lee, G. H.; Pearton, S. J. Two-Dimensional Material Templates for van Der Waals Epitaxy, Remote Epitaxy, and Intercalation Growth. *Appl Phys Rev* **2022**, *9* (3), 031305. https://doi.org/10.1063/5.0090373.

(20) In Yoon, S.; Park, H.; Lee, Y.; Guo, C.; Jin Kim, Y.; Song Lee, J.; Son, S.; Choe, M.; Han, D.; Kwon, K.; Lee, J.; Yeol Ma, K.; Ghassami, A.; Wook Moon, S.; Park, S.-Y.; Kyun Kang, B.; Kim, Y.-J.; Koo, S.; Genco, A.; Shim, J.; Tartakovskii, A.; Duan, Y.; Ding, F.; Ahn, S.; Ryu, S.; Kim, J.-Y.; Seok Yang, W.; Chhowalla, M.; Park, Y. S.; Kyu Min, S.; Lee, Z.; Suk Shin, H. Pressure Enabled Organic Reactions via Confinement between Layers of 2D Materials. *Sci. Adv* **2024**, *10* (45), eadp9804. https://doi.org/10.1126/sciadv.adp9804.

(21) Deng, D.; Novoselov, K. S.; Fu, Q.; Zheng, N.; Tian, Z.; Bao, X. Catalysis with Two-Dimensional Materials and Their Heterostructures. *Nat Nanotechnol* **2016**, *11* (3), 218–230. https://doi.org/10.1038/nnano.2015.340.

(22) Shifa, T. A.; Vomiero, A. Confined Catalysis: Progress and Prospects in Energy Conversion. *Adv Energy Mater* **2019**, *9* (40), 1902307. https://doi.org/10.1002/aenm.201902307.

(23) Shih, A. J.; Arulmozhi, N.; Koper, M. T. M. Electrocatalysis under Cover: Enhanced Hydrogen Evolution via Defective Graphene-Covered Pt(111). *ACS Catal* **2021**, *11* (17), 10892–10901. https://doi.org/10.1021/acscatal.1c02145.

(24) Sutter, P.; Sadowski, J. T.; Sutter, E. A. Chemistry under Cover: Tuning Metal-Graphene Interaction by Reactive Intercalation. *J Am Chem Soc* **2010**, *132* (23), 8175–8179. https://doi.org/10.1021/ja102398n.

(25) Fu, Q.; Bao, X. Surface Chemistry and Catalysis Confined under Two-Dimensional Materials. *Chem Soc Rev* **2017**, *46* (7), 1842–1874. https://doi.org/10.1039/c6cs00424e.

(26) Zhang, L.; Ng, M. L.; Vojvodic, A. Role of Undercoordinated Sites for the Catalysis in Confined Spaces Formed by Two-Dimensional Material Overlayers. *Journal of Physical Chemistry Letters* **2020**, *11* (21), 9400–9407. https://doi.org/10.1021/acs.jpclett.0c02652.

(27) Grånäs, E.; Schröder, U. A.; Arman, M. A.; Andersen, M.; Gerber, T.; Schulte, K.; Andersen, J. N.; Michely, T.; Hammer, B.; Knudsen, J. Water Chemistry beneath Graphene: Condensation of a Dense OH-H2O Phase under Graphene. *Journal of Physical Chemistry C* **2022**, *126* (9), 4347–4354. https://doi.org/10.1021/acs.jpcc.1c10289.

(28) Wei, F.; Wan, Q.; Lin, S.; Guo, H. Origin of Confined Catalysis in Nanoscale Reactors between Two-Dimensional Covers and Metal Substrates: Mechanical or Electronic? *Journal of Physical Chemistry C* **2020**, *124* (21), 11564–11573. https://doi.org/10.1021/acs.jpcc.0c03621.

(29) Wang, W. X.; Wei, Y. W.; Li, S. Y.; Li, X.; Wu, X.; Feng, J.; He, L. Imaging the Dynamics of an Individual Hydrogen Atom Intercalated between Two Graphene Sheets. *Phys Rev B* **2018**, *97* (8), 085407. https://doi.org/10.1103/PhysRevB.97.085407.

(30) Yao, Y.; Fu, Q.; Zhang, Y. Y.; Weng, X.; Li, H.; Chend, M.; Jin, L.; Dong, A.; Mu, R.; Jiang, P.; Liu, L.; Bluhm, H.; Liu, Z.; Zhang, S. B.; Bao, X. Graphene Cover-Promoted Metal-Catalyzed Reactions. *Proc Natl Acad Sci U S A* **2014**, *111* (48), 17023–17028. https://doi.org/10.1073/pnas.1416368111.

(31) Boix, V.; Scardamaglia, M.; Gallo, T.; D'Acunto, G.; Strømsheim, M. D.; Cavalca, F.; Zhu, S.; Shavorskiy, A.; Schnadt, J.; Knudsen, J. Following the Kinetics of Undercover Catalysis with APXPS and the Role of Hydrogen as an Intercalation Promoter. *ACS Catal* **2022**, *12* (16), 9897–9907. https://doi.org/10.1021/acscatal.2c00803.

(32) Starodub, E.; Bartelt, N. C.; Mccarty, K. F. Oxidation of Graphene on Metals. *J. Phys. Chem. C* **2010**, *1* (1), 5134–5140. https://doi.org/https://doi.org/10.1021/jp912139e.

(33) Grånäs, E.; Knudsen, J.; Schröder, U. A.; Gerber, T.; Busse, C.; Arman, M. A.; Schulte, K.; Andersen, J. N.; Michely, T. Oxygen Intercalation under Graphene on Ir(111): Energetics, Kinetics, and the Role of Graphene Edges. *ACS Nano* **2012**, *6* (11), 9951–9963. https://doi.org/10.1021/nn303548z.

(34) Ma, L.; Zeng, X. C.; Wang, J. Oxygen Intercalation of Graphene on Transition Metal Substrate: An Edge-Limited Mechanism. *Journal of Physical Chemistry Letters* **2015**, *6* (20), 4099–4105. https://doi.org/10.1021/acs.jpclett.5b01841.

(35) Zhang, X.; Wang, L.; Xin, J.; Yakobson, B. I.; Ding, F. Role of Hydrogen in Graphene Chemical Vapor Deposition Growth on a Copper Surface. *J Am Chem Soc* **2014**, *136* (8), 3040–3047. https://doi.org/10.1021/ja405499x.

(36) Weatherup, R. S.; D'Arsié, L.; Cabrero-Vilatela, A.; Caneva, S.; Blume, R.; Robertson, J.; Schloegl, R.; Hofmann, S. Long-Term Passivation of Strongly Interacting Metals with Single-Layer Graphene. *J Am Chem Soc* **2015**, *137* (45), 14358–14366. https://doi.org/10.1021/jacs.5b08729.

(37) Kühne, M.; Paolucci, F.; Popovic, J.; Ostrovsky, P. M.; Maier, J.; Smet, J. H. Ultrafast Lithium Diffusion in Bilayer Graphene. *Nat Nanotechnol* **2017**, *12* (9), 895–900. https://doi.org/10.1038/nnano.2017.108.

(38) Li, W.; Yi, D. Adatom Diffusion beneath Graphene. *J. Phys. Chem. C* **2025**, *129* (8), 4165–4171. https://doi.org/10.1021/acs.jpcc.4c06865.

(39) Kapustić, K.; G. Ayani, C.; Pielić, B.; Plevová, K.; Mandić, Š.; Šrut Rakić, I. Visualizing Intercalation Effects in 2D Materials Using AFM-Based Techniques. *Journal of Physical Chemistry Letters* **2025**, *16* (19), 4804–4811. https://doi.org/10.1021/acs.jpclett.5c00322.

(40) Wang, Z. J.; Liang, Z.; Kong, X.; Zhang, X.; Qiao, R.; Wang, J.; Zhang, S.; Zhang, Z.; Xue, C.; Cui, G.; Zhang, Z.; Zou, D.; Liu, Z.; Li, Q.; Wei, W.; Zhou, X.; Tang, Z.; Yu, D.; Wang, E.; Liu, K.; Ding, F.; Xu, X. Visualizing the Anomalous Catalysis in Two-Dimensional Confined Space. *Nano Lett* **2022**, *22* (12), 4661–4668. https://doi.org/10.1021/acs.nanolett.2c00549.

(41) Grånäs, E.; Gerber, T.; Schröder, U. A.; Schulte, K.; Andersen, J. N.; Michely, T.; Knudsen, J. Hydrogen Intercalation under Graphene on Ir(111). *Surf Sci* **2016**, *651*, 57–61. https://doi.org/10.1016/j.susc.2016.03.002.

(42) Jin, L.; Fu, Q.; Mu, R.; Tan, D.; Bao, X. Pb Intercalation underneath a Graphene Layer on Ru(0001) and Its Effect on Graphene Oxidation. *Physical Chemistry Chemical Physics* **2011**, *13* (37), 16655–16660. https://doi.org/10.1039/c1cp21843c.

(43) Mu, R.; Fu, Q.; Jin, L.; Yu, L.; Fang, G.; Tan, D.; Bao, X. Visualizing Chemical Reactions Confined under Graphene. *Angewandte Chemie - International Edition* **2012**, *51* (20), 4856–4859. https://doi.org/10.1002/anie.201200413.

(44) Steininger, H.; Lehwald, S.; Ibach, H. Adsorption of Oxygen on Pt(111). *Surf Sci* **1982**, *123* (1), 1–17. https://doi.org/10.1016/0039-6028(82)90124-8.

(45) Steininger, H.; Lehwald, S.; Ibach, H. On the Adsorption of CO on Pt(111). *Surf Sci* **1982**, *123* (2–3), 264–282. https://doi.org/https://doi.org/10.1016/0039-6028(82)90328-4.

(46) Ogletree, D. F.; Van Hove, M. A.; Somorjai, G. A. LEED Intensity Analysis of the Structures of Clean Pt(111) and of CO Adsorbed on Pt(111) in the c(4 × 2) Arrangement. *Surf Sci* **1986**, *173* (2–3), 351–365. https://doi.org/10.1016/0039-6028(86)90195-0.

(47) Graïšnäs, E.; Andersen, M.; Arman, M. A.; Gerber, T.; Hammer, B.; Schnadt, J.; Andersen, J. N.; Michely, T.; Knudsen, J. CO Intercalation of Graphene on Ir(111) in the Millibar Regime. *Journal of Physical Chemistry C* **2013**, *117* (32), 16438–16447. https://doi.org/10.1021/jp4043045.

(48) Murata, Y.; Starodub, E.; Kappes, B. B.; Ciobanu, C. V.; Bartelt, N. C.; McCarty, K. F.; Kodambaka, S. Orientation-Dependent Work Function of Graphene on Pd(111). *Appl Phys Lett* **2010**, *97* (14), 143114. https://doi.org/10.1063/1.3495784.

(49) Kaack, M.; Fick, D. Determination of the Work Functions of Pt(111) and Ir(111) beyond 1100 K Surface Temperature. *Surf Sci* **1995**, *342* (1–3), 111–118. https://doi.org/10.1016/0039-6028(95)00758-X.

(50) Srivastava, N.; Gao, Q.; Widom, M.; Feenstra, R. M.; Nie, S.; McCarty, K. F.; Vlassiouk, I. V. Low-Energy Electron Reflectivity of Graphene on Copper and Other Substrates. *Phys Rev B* **2013**, *87* (24), 245414. https://doi.org/10.1103/PhysRevB.87.245414.

(51) Wintterlin, J.; Schuster, R.; Ertl, G. Existence of a "Hot" Atom Mechanism for the Dissociation of O 2 on Pt(111). *Phys. Rev. Lett.* **1996**, *77* (123), 123–126. https://doi.org/10.1103/PhysRevLett.77.123.

(52) Von Oertzen, A.; Rotermund, H. H.; Nettesheim, S. Diffusion of Carbon Monoxide and Oxygen on Pt( 110)" Experiments Performed with the PEEM. *Surface Science* **1994**, *311* (3), 322–330. https://doi.org/10.1016/0039-6028(94)91422-2.

(53) Seebauer, E. G.; Schmidt, L. D. Surface Diffusion of Hydrogen on Pt(111): Laser-Induced Thermal Desorption Studies. *Chem Phys Lett* **1986**, *123* (1–2), 129–133. https://doi.org/10.1016/0009-2614(86)87027-0.

(54) Li, T.; Yarmoff, J. A. Intercalation and Desorption of Oxygen between Graphene and Ru(0001) Studied with Helium Ion Scattering. *Phys Rev B* **2017**, *96* (15), 155441. https://doi.org/10.1103/PhysRevB.96.155441.

(55) Grebenko, A. K.; Krasnikov, D. V.; Bubis, A. V.; Stolyarov, V. S.; Vyalikh, D. V.; Makarova, A. A.; Fedorov, A.; Aitkulova, A.; Alekseeva, A. A.; Gilshtein, E.; Bedran, Z.; Shmakov, A. N.; Alyabyeva, L.; Mozhchil, R. N.; Ionov, A. M.; Gorshunov, B. P.; Laasonen, K.; Podzorov, V.; Nasibulin, A. G. High-Quality Graphene Using Boudouard Reaction. *Advanced Science* **2022**, *9* (12), 2200217. https://doi.org/10.1002/advs.202200217.

(56) McCrea, K. R.; Parker, J. S.; Somorjai, G. A. The Role of Carbon Deposition from CO Dissociation on Platinum Crystal Surfaces during Catalytic CO Oxidation: Effects on Turnover Rate, Ignition Temperature, and Vibrational Spectra. *Journal of Physical Chemistry B* **2002**, *106* (42), 10854–10863. https://doi.org/10.1021/jp014679k.

(57) Jin, L.; Zhao, C.; Gong, Z.; Pan, J.; Wei, W.; Wang, G.; Cui, Y. Hydrogen-Promoted Graphene Growth on Pt(111) via CVD Methods. *Surfaces and Interfaces* **2021**, *26*, 101383. https://doi.org/10.1016/j.surfin.2021.101383.

(58) Prieto, M. J.; Klemm, H. W.; Xiong, F.; Gottlob, D. M.; Menzel, D.; Schmidt, T.; Freund, H. J. Water Formation under Silica Thin Films: Real-Time Observation of a Chemical Reaction in a Physically Confined Space. *Angewandte Chemie - International Edition* **2018**, *57* (28), 8749–8753. https://doi.org/10.1002/anie.201802000.

(59) Wang, M.; Wang, M.; Zhou, C.; Zhou, C.; Akter, N.; Akter, N.; Tysoe, W. T.; Boscoboinik, J. A.; Lu, D. Mechanism of the Accelerated Water Formation Reaction under Interfacial Confinement. *ACS Catal* **2020**, *10* (11), 6119–6128. https://doi.org/10.1021/acscatal.9b05289.

(60) Reyerson, L. H.; Kobe, K. Carbon Suboxide. **1930**, *7* (4), 479–492. https://doi.org/10.1021/cr60028a002.

(61) Strudwick, A. J.; Weber, N. E.; Schwab, M. G.; Kettner, M.; Weitz, R. T.; Wünsch, J. R.; Müllen, K.; Sachdev, H. Chemical Vapor Deposition of High Quality Graphene Films from Carbon Dioxide Atmospheres. *ACS Nano* **2015**, *9* (1), 31–42. https://doi.org/10.1021/nn504822m.

(62) Yang, X.; Hu, B.; Jin, Y.; Zhao, W.; Luo, Z.; Lu, Z.; Fang, L.; Ruan, H. Insight into $CO_2$ Etching Behavior for Efficiently Nanosizing Graphene. *Adv Mater Interfaces* **2017**, *4* (10), 1601065. https://doi.org/10.1002/admi.201601065.

(63) Plimpton, S. Fast Parallel Algorithms for Short-Range Molecular Dynamics. *J Comput Phys* **1995**, *117* (1), 1–19. https://doi.org/10.1006/jcph.1995.1039.

(64) Singh, S. K.; Srinivasan, S. G.; Neek-Amal, M.; Costamagna, S.; Van Duin, A. C. T.; Peeters, F. M. Thermal Properties of Fluorinated Graphene. *Phys Rev B Condens Matter Mater Phys* **2013**, *87* (10), 104114. https://doi.org/10.1103/PhysRevB.87.104114.

(65) Nosé, S. A Unified Formulation of the Constant Temperature Molecular Dynamics Methods. *J Chem Phys* **1984**, *81* (1), 511–519. https://doi.org/10.1063/1.447334.

(66) Swope, W. C.; Andersen, H. C.; Berens, P. H.; Wilson, K. R. A Computer Simulation Method for the Calculation of Equilibrium Constants for the Formation of Physical Clusters of Molecules: Application to Small Water Clusters. *J Chem Phys* **1982**, *76* (1), 637–649. https://doi.org/10.1063/1.442716.

**Acknowledgments.** This research was supported by Quantum materials for applications in sustainable technologies (QM4ST) - project No. CZ.02.01.01/00/22_008/0004572 by OP JAK, call Excellent Research and Ministry of Education, Youth and Sports of the Czech Republic - LM2023051.

**Competing interests**: The authors have no competing interests to declare.

**Data and materials availability**: The data underlying this study are openly available at https://doi.org/10.5281/zenodo.15849416.

# Mass-transport-limited reaction rates and molecular diffusion in the van der Waals gap beneath graphene

Hossein Mirdamadi[1#], Jiří David[1,2#], Rui Wang[3], Tianle Jiang[4], Yanming Wang[5], Karel Vařeka[2], Michal Dymáček[2], Petr Bábor[1,2], Tomáš Šikola[1,2], Miroslav Kolíbal[1,2*]

[1]CEITEC BUT, Brno University of Technology, Purkyňova 123, 612 00 Brno, Czech Republic

[2]Institute of Physical Engineering, Brno University of Technology, Technická 2, 616 69 Brno, Czech Republic

[3]School of Mechanical Engineering, Shanghai Jiao Tong University, 800 Dongchuan Rd, Shanghai, China, 200240

[4]University of Michigan-Shanghai Jiao Tong University Joint Institute, Shanghai Jiao Tong University, 800 Dongchuan Rd, Shanghai, China, 200240

[5]Global Institute of Future Technology, Shanghai Jiao Tong University, 800 Dongchuan Rd, Shanghai, China, 200240

[#]contributed equally to this work

[*]kolibal.m@fme.vutbr.cz

Contains supporting figures S1-S10.

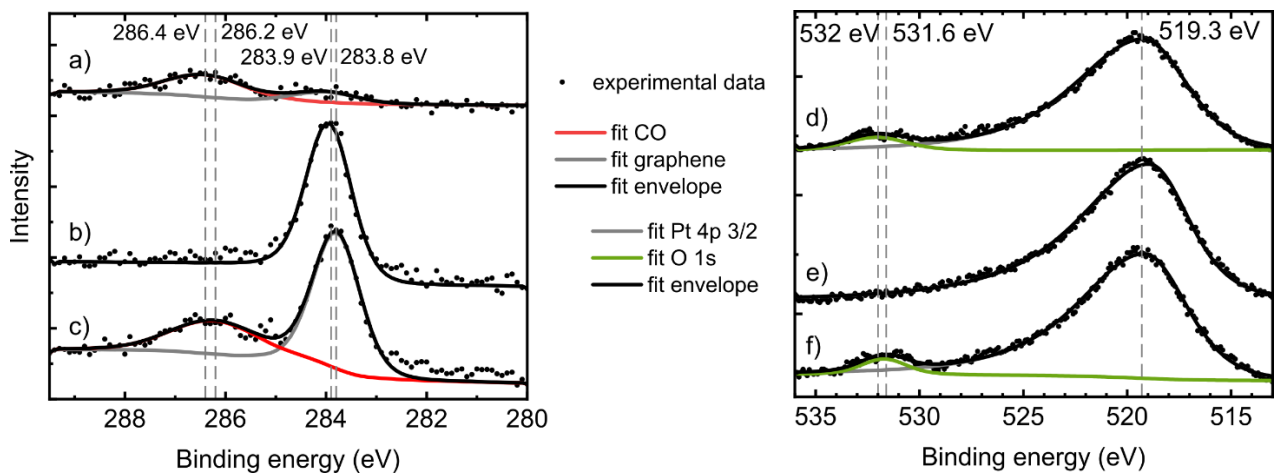

Fig. S1: XPS analysis of (a,d) CO-adsorbed Pt, (b,e) non-intercalated graphene/Pt and (c,f) CO-intercalated graphene/Pt. Left: C 1s area, right: O 1s area.

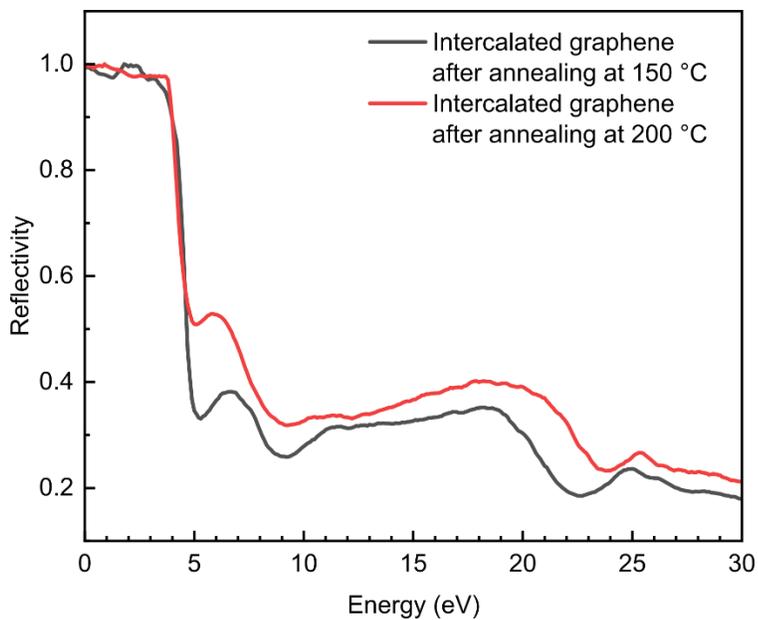

Fig. S2: Electron reflectivity of CO-intercalated graphene at elevated temperatures. The two minima between 5 and 10 eV persist.

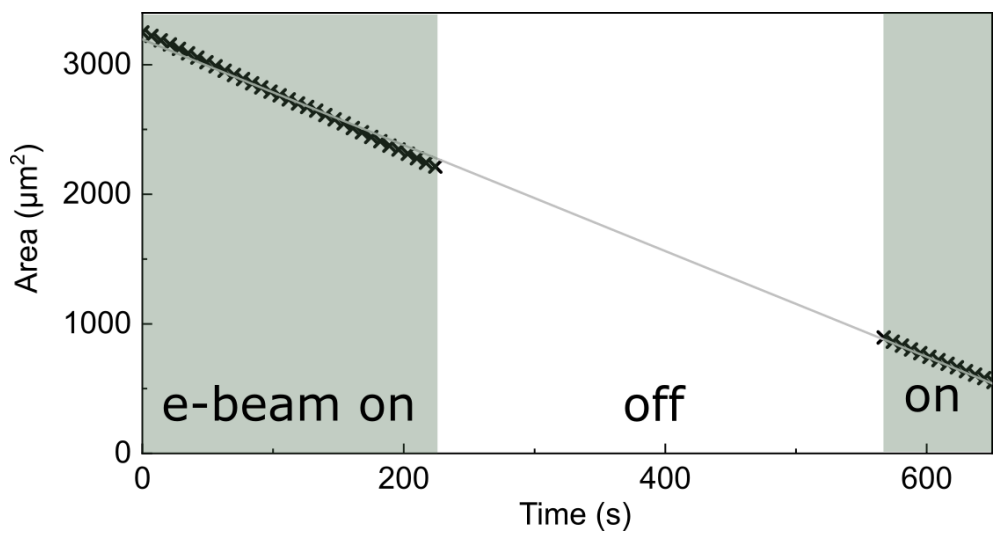

Fig. S3: The effect of the electron beam exposure (5 keV, 300 pA) on the observed graphene area reduction during etching by oxygen gas (T = 1000 °C, p = 1.6 × 10$^{-6}$ Pa). The deviation from the grey line (which represents constant etch rate) is negligible.

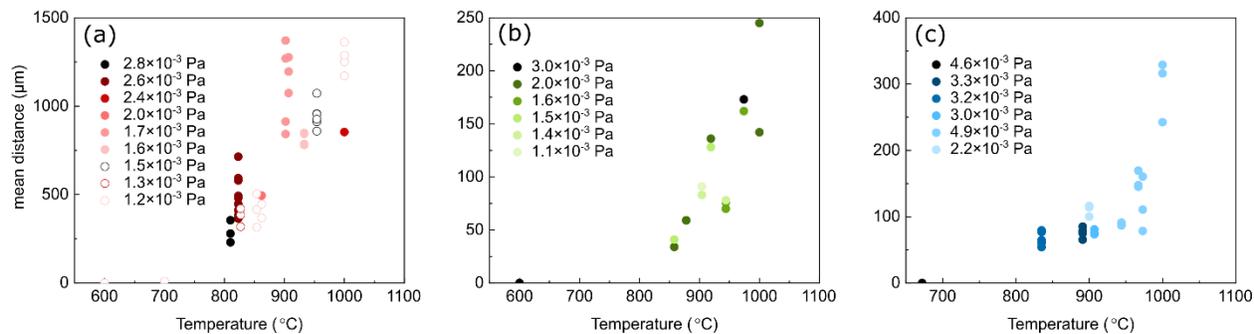

Fig. S4: The mean distance (between the overlayer etch front and the edge of buried graphene) dependence on temperature, as measured for distinct pressures (as marked in the legends) and different gases: (a) CO, (b) $O_2$ (c) $H_2$.

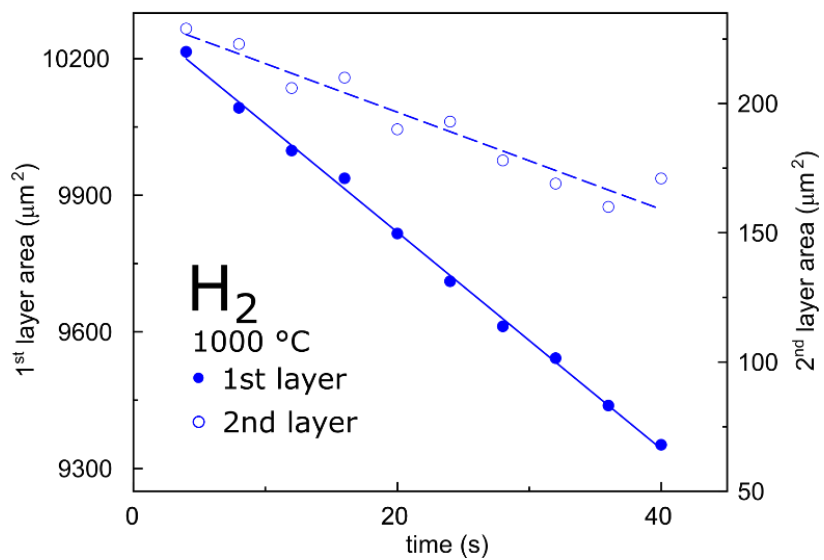

Fig. S5: Area reduction of first (overlayer) and second (buried) graphene layer versus time during exposure to hydrogen at 1000 °C (7.44 × 10$^{-4}$ Pa).

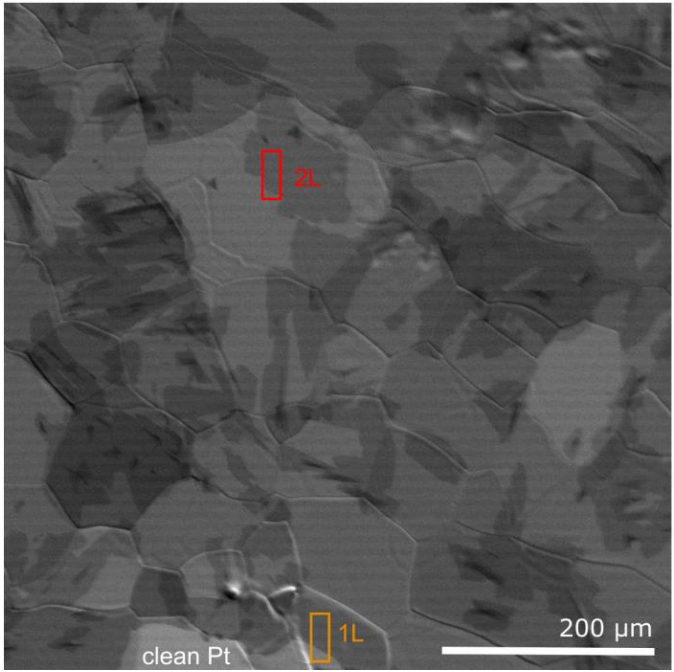

Fig. S6: Demonstration of etch rate data evaluation. We have derived the etch rates by segmenting the graphene area reduction in the denoted rectangles (orange: graphene overlayer, red: buried graphene layer). Note that the inspected graphene areas are localized atop different platinum grains, which could affect the etch rate. Nevertheless, in all cases we have observed the overlayer being etched faster than buried graphene layers.

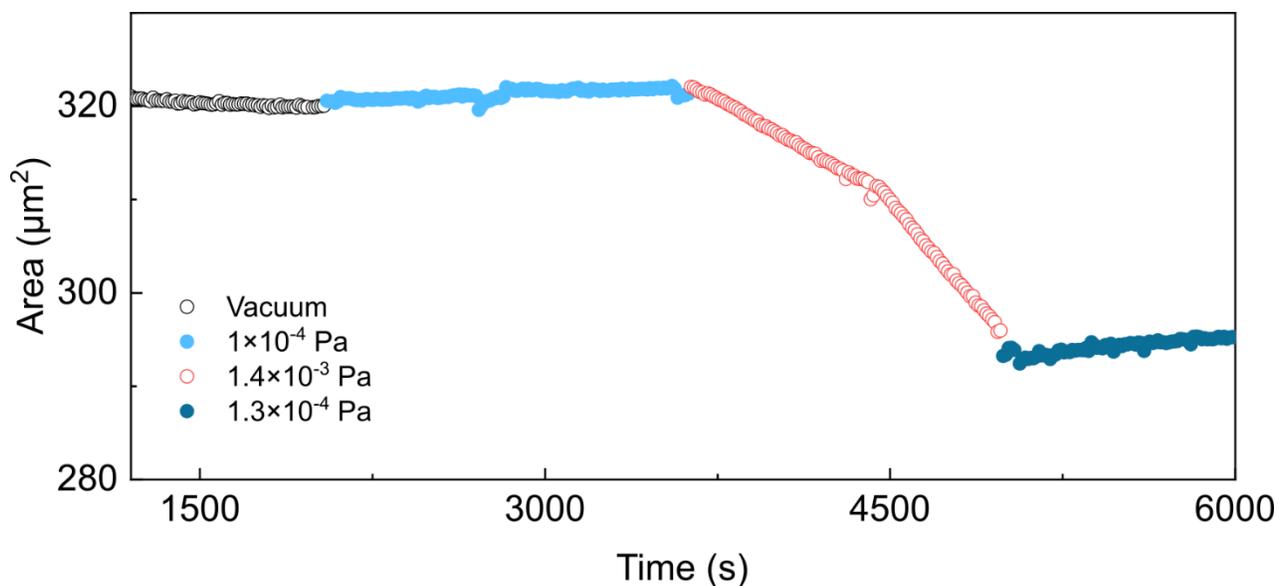

Fig. S7: Stability of graphene at low CO pressures. Time-dependence of graphene flake area (as calculated from SEM images) at 834 °C at different CO pressures. At low partial CO pressures (low $10^{-4}$ Pa range) the flake area slightly increases, possibly due to Boudouard reaction, $2CO \leftrightarrow C+CO_2$, similarly to the other reports [55]. If the CO pressure is increased up to $10^{-3}$ Pa range, the flake behaviour is reversed and instead of growth we observe rapid etching of the flake.

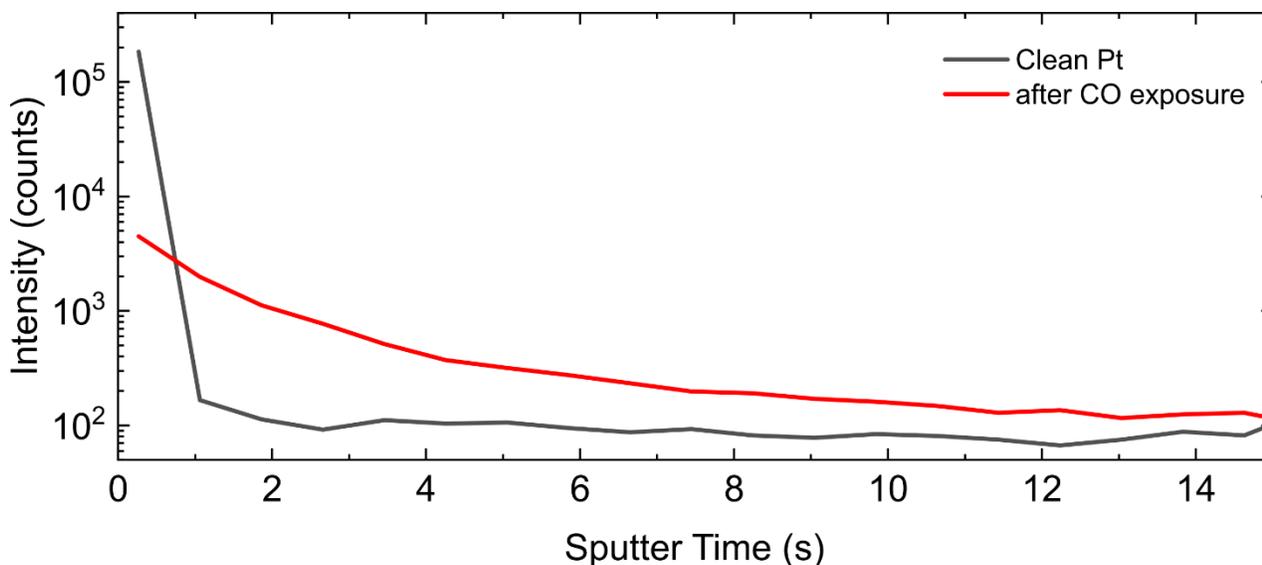

Fig. S8: SIMS depth profile (C$^-$ signal) of platinum exposed to CO. Grey data: depth profile of clean platinum surface (annealed in oxygen, 1.2×10$^{-4}$ Pa, for 10 minutes at temperature slightly above 1000 °C). Red data: depth profile of the same sample after exposure to CO (1.8×10$^{-3}$ Pa for 40 minutes at temperature slightly above 1000 °C. We first evacuated all the CO from the chamber and only then cooled down the sample to room temperature, to avoid false carbon-related signal detection, potentially due to adsorbed CO molecules.

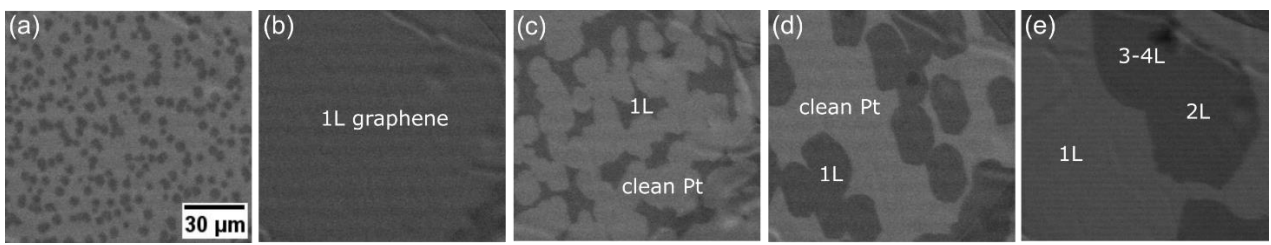

Fig. S9: SEM sequence demonstrating the wedding cake graphene formation. (a,b) A full graphene overlayer (1L) is nucleated on Pt at 1000 °C by exposure to ethylene. Then, the ethylene flow is stopped. (c) As the temperature is increased to 1200 °C, graphene gets dissolved in Pt substrate. By decreasing the temperature below 1000 °C, the dissolved carbon segregates at the Pt surface and forms graphene overlayer (d). Subsequently, multilayer graphene is formed by additional carbon segregation below the graphene overlayer (e).

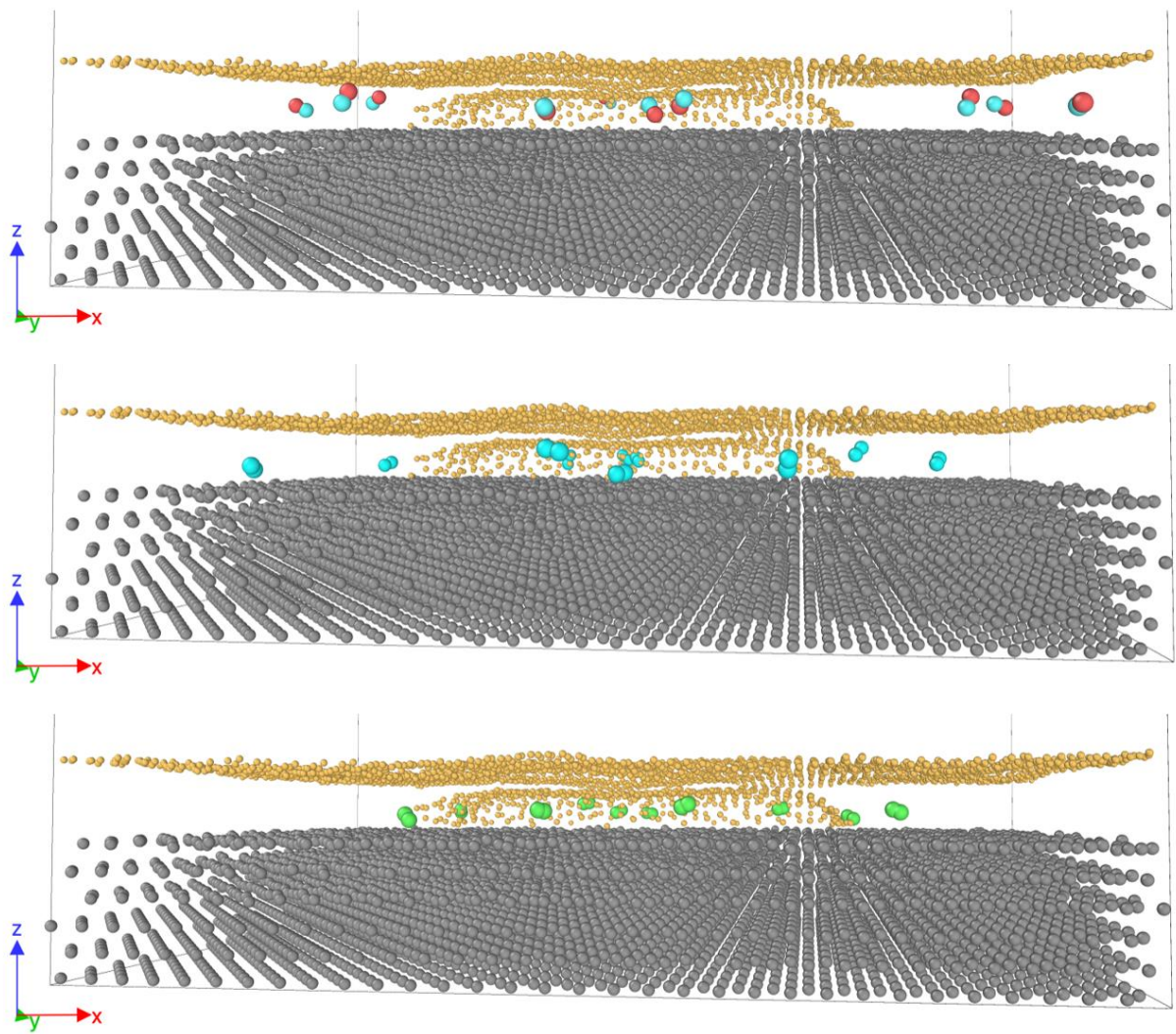

Fig. S10: Initial MD configuration of the intercalated graphene/Pt system. Gray: Pt atoms. Orange: graphene. Red: C atom in CO molecules. Blue: O atoms. Green: H atoms.